\documentclass[12pt,a4paper]{article}
\usepackage{a4wide}
\usepackage{amsmath}
\usepackage{amssymb}
\usepackage{amsfonts}
\usepackage{epsfig}
\usepackage{subfigure}
\usepackage{exscale}
\usepackage{float}
\usepackage{bbm}
\usepackage[numbers,sort&compress]{natbib}
\newcommand{\N}{{\mathbb{N}}}
\newcommand{\Z}{{\mathbb{Z}}}
\newcommand{\R}{{\mathbb{R}}}
\newcommand{\C}{{\mathbb{C}}}

\newcommand{\p}{\partial}
\newcommand{\bra}{\langle}
\newcommand{\ket}{\rangle}

\makeatletter
\@addtoreset{equation}{section}
\makeatother

\title{Runge-Lenz Vector, Accidental $SU(2)$ Symmetry, \\
and Unusual Multiplets for Motion on a Cone}

\author{M.\ H.\ Al-Hashimi and U.-J.\ Wiese \\ \\
Institute for Theoretical Physics, Bern University \\
Sidlerstrasse 5, CH-3012 Bern, Switzerland \\ \\}

\begin{document} 

\maketitle

\vspace{-1cm}

\begin{abstract} \normalsize

We consider a particle moving on a cone and bound to its tip by $1/r$ or 
harmonic oscillator potentials. When the deficit angle of the cone divided by 
$2 \pi$ is a rational number, all bound classical orbits are closed. 
Correspondingly, the quantum system has accidental degeneracies in the discrete
energy spectrum. An accidental $SU(2)$ symmetry is generated by the rotations 
around the tip of the cone as well as by a Runge-Lenz vector. Remarkably, some 
of the corresponding multiplets have fractional ``spin'' and unusual 
degeneracies. 

\end{abstract}
 
\section{Introduction}

It is well-known that $1/r$ and harmonic oscillator potentials are exceptional
because, in addition to rotation invariance, they have accidental dynamical 
symmetries. At the classical level the accidental symmetries imply that all 
bound orbits are closed, while at the quantum level they give rise to 
additional degeneracies in the discrete energy spectrum. In particular, the 
$SO(d)$ rotational symmetry of the $d$-dimensional $1/r$ potential (the Coulomb
potential for $d = 3$) is enlarged to the accidental symmetry $SO(d+1)$. The 
additional conserved quantities form the components of the Runge-Lenz vector. 
Similarly, the $d$-dimensional harmonic oscillator has an $SO(d)$ rotational 
symmetry which is contained as a subgroup in an accidental $SU(d)$ symmetry.

It has been shown by Bertrand in 1873 that the $1/r$ and $r^2$ potentials are
the only spherically symmetric scalar potentials in Euclidean space for which 
all bound orbits are 
closed \cite{Ber73}. Still, there exist a number of other systems with 
accidental symmetries involving vector potentials or non-Euclidean spaces. For
example, a free particle confined to the surface of the $d$-dimensional 
hyper-sphere $S^d$ moves along a great circle (which obviously is closed). 
Indeed the rotational $SO(d+1)$ symmetry of this system corresponds to the 
accidental symmetry of the $1/r$ potential. Already in 1935 Fock has realized 
that the hydrogen atom possesses ``hyper-spherical'' symmetry \cite{Foc35}. 
Based on this work, Bargmann \cite{Bar36} has shown that the  generators of the
accidental symmetry are the components of the Runge-Lenz vector \cite{Len24}
\begin{equation}
\label{RungeLenz}
\vec R = 
\frac{1}{2 M} \left(\vec p \times \vec L - \vec L \times \vec p\right) -
\kappa \vec e_r.
\end{equation} 
Here $\vec p$ and $\vec L$ are the momentum and angular momentum of a particle 
of mass $M$, $\kappa$ is the strength of the $1/r$ potential, and $\vec e_r$ 
is the radial unit-vector. An example of an accidental symmetry involving a 
vector potential is cyclotron motion \cite{Lan30,Joh49}. In all these cases, 
there is a deep connection between the fact that all bound classical orbits are
closed and additional degeneracies in the discrete energy spectrum of the
corresponding quantum system. The subject of accidental symmetry has been 
reviewed, for example, by McIntosh \cite{McI71}.

In order to further investigate the phenomenon of accidental symmetries, in 
this paper we study a particle confined to the surface of a cone. A cone is 
obtained from the plane by removing a wedge of deficit angle $\delta$ and 
gluing the open ends back together. As a consequence, the polar angle $\chi$ no
longer extends from $0$ to $2 \pi$, but only to $2 \pi - \delta$. The geometry 
of the cone is illustrated in figure 1.
\begin{figure}[htb]
\begin{center}
\epsfig{file=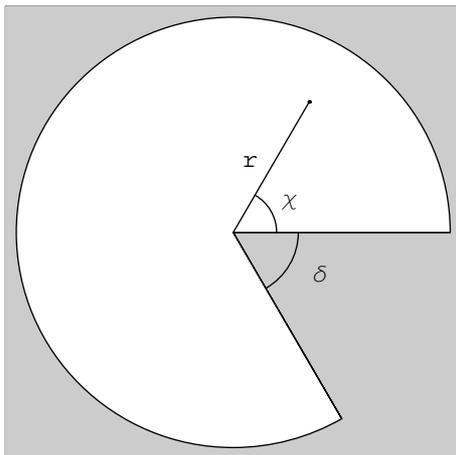,width=6cm}
\end{center}
\caption{\it A cone is obtained by cutting a wedge of deficit angle $\delta$ 
out of the 2-dimensional plane, and by gluing the open ends back together. 
Points on the cone are described by the distance $r$ from the tip and an angle 
$\chi$ which varies between $0$ and $2 \pi - \delta$. Unlike the cone in the 
figure, the actual cone considered in this work extends over the whole range 
$r \in (0,\infty)$.}
\end{figure}
It is convenient to rescale the polar angle such that it again covers the full 
interval, i.e.\
\begin{equation}
\varphi = \frac{\chi}{s} \in [0,2 \pi],
\end{equation}
with the scale factor
\begin{equation}
s =  1 - \frac{\delta}{2 \pi}.
\end{equation}
The kinetic energy of a particle of mass $M$ then takes the form
\begin{equation}
T = \frac{M}{2} (\dot{r}^2 + r^2 \dot{\chi}^2) = 
\frac{M}{2} (\dot{r}^2 + r^2 s^2 \dot{\varphi}^2).
\end{equation}
The radial component of the momentum $p_r$ is canonically conjugate to $r$, 
i.e.
\begin{equation}
p_r = \frac{\p T}{\p \dot{r}} = M \dot{r}.
\end{equation}
Similarly, the canonically conjugate momentum corresponding to the rescaled 
angle $\varphi$ is given by the angular momentum
\begin{equation}
L = \frac{\p T}{\p \dot{\varphi}} = M r^2 s^2 \dot{\varphi},
\end{equation}
such that
\begin{equation}
T = \frac{1}{2M} \left(p_r^2 + \frac{L^2}{r^2 s^2}\right).
\end{equation}

As usual, upon canonical quantization (and using natural units in which 
$\hbar = 1$) the angular momentum conjugate to the rescaled angle $\varphi$ is 
represented by the operator
\begin{equation}
L = - i \p_\varphi.
\end{equation}
The issues of domains of operators in a corresponding Hilbert space as well as 
of Hermiticity and self-adjointness play a certain role in this paper. For some
mathematical background we refer to \cite{Ree72,Wei87}. Let us 
begin to address these issues in the context of the operator $L$. First of all,
the Hilbert space ${\cal H} = L_2((0,\infty) \times (0,2 \pi);r)$ for a 
particle moving on a cone consists of the square-integrable functions 
$\Psi(r,\varphi)$ with $r \in (0,\infty)$, $\varphi \in [0,2 \pi]$, and with 
the norm $\langle\Psi|\Psi\rangle < \infty$, which is induced by the scalar 
product 
\begin{equation}
\langle\Phi|\Psi\rangle = \int_0^\infty dr \ r \int_0^{2 \pi} d\varphi \ 
\Phi(r,\varphi)^* \Psi(r,\varphi).
\end{equation}
It should be noted that functions in the Hilbert space need not be continuous 
or differentiable, in particular, they need not be periodic. In order to 
completely define a quantum mechanical operator $O$, one must identify the 
domain ${\cal D}[O] \subset {\cal H}$ of wave functions on which the operator
acts. Since the self-adjoint operator $L$ is unbounded in ${\cal H}$, its 
domain is dense in ${\cal H}$ but not equal to ${\cal H}$. In order to be able
to act on it with $L$, a wave function $\Psi$ must be differentiable at least 
once with respect to $\varphi$.

An operator $O$ is said to be Hermitean (symmetric in mathematical parlance) if
\begin{equation}
\langle O \Phi|\Psi \rangle = \langle \Phi|O \Psi\rangle,
\end{equation}
for all wave functions $\Phi, \Psi \in {\cal D}[O]$. In order to investigate 
the question of Her\-mi\-ti\-ci\-ty of $L$, we perform a partial integration
and obtain
\begin{eqnarray}
\langle O \Phi|\Psi \rangle&=&\int_0^\infty dr \ r \int_0^{2 \pi} d\varphi \ 
[- i \p_\varphi \Phi(r,\varphi)]^* \Psi(r,\varphi) \nonumber \\
&=&\int_0^\infty dr \ r \int_0^{2 \pi} d\varphi \ 
\Phi(r,\varphi)^* [-i \p_\varphi \Psi(r,\varphi)] \nonumber \\
&+&i \int_0^\infty dr \ r 
\Phi(r,\varphi)^* \Psi(r,\varphi)|_{\varphi = 0}^{\varphi = 2 \pi} \nonumber \\
&=&\langle \Phi|O \Psi\rangle + i \int_0^\infty dr \ r 
\Phi(r,\varphi)^* \Psi(r,\varphi)|_{\varphi = 0}^{\varphi = 2 \pi}.
\end{eqnarray}
Thus, the operator $L$ is Hermitean if
\begin{equation}
\label{condition}
\Phi(r,\varphi)^* \Psi(r,\varphi)|_{\varphi = 0}^{\varphi = 2 \pi} = 0.
\end{equation}

Let us now address the issue of self-adjointness versus Hermiticity. In 
particular, self-adjointness (but not Hermiticity alone) guarantees a 
real-valued spectrum. An operator $O$ is self-adjoint (i.e.\ $O = O^\dagger$)
if it is Hermitean and the domain of its adjoint $O^\dagger$ coincides with the
domain of $O$, i.e.\ ${\cal D}[O^\dagger] = {\cal D}[O]$. The domain 
${\cal D}[O^\dagger]$ consists of all functions $\xi \in {\cal H}$ for which 
there exists a function $\eta \in  {\cal H}$ such that
\begin{equation}
\langle \eta |\Psi\rangle = \langle \xi|O \Psi\rangle,
\end{equation}
for all $\Psi \in {\cal D}[O]$. For $\xi \in {\cal D}[O^\dagger]$ one then has
$O^\dagger \xi = \eta$.
For example, let us consider the operator $L$ in the domain of differentiable 
functions $\Psi \subset {\cal H}$ (with $L \Psi \in {\cal H}$), which obey the 
boundary condition $\Psi(r,0) = \Psi(r,2 \pi) = 0$. In this domain $L$ acts as 
a Hermitean operator because the condition of eq.(\ref{condition}) is indeed
satisfied. However, the functions in the domain of $L^\dagger$ need not obey 
this condition (i.e.\ the domain of $L^\dagger$ is larger than the one of $L$) 
and thus $L$ restricted to the above domain is not self-adjoint. However, there
is a family of self-adjoint extensions. To see this, let us extend the domain
of $L$ to differentiable functions obeying
\begin{equation}
\Psi(r,2 \pi) = z \Psi(r,0), \ z \in \C. 
\end{equation}
Then the condition of eq.(\ref{condition}) implies
\begin{equation}
\Phi(r,2 \pi)^* \Psi(r,2 \pi) - \Phi(r,0)^* \Psi(r,0) = 
[\Phi(r,2 \pi)^* z - \Phi(r,0)^*] \Psi(r,0) = 0,
\end{equation}
such that
\begin{equation}
\Phi(r,2 \pi) = \frac{1}{z^*} \Phi(r,0). 
\end{equation}
In order to be self-adjoint (i.e.\ to have ${\cal D}[L^\dagger] = {\cal D}[L]$)
the functions $\Phi \in {\cal D}[L^\dagger]$ must obey the same condition as 
$\Psi \in {\cal D}[L]$, which implies $z = 1/z^* = \exp(i \theta)$. The angle 
$\theta$ characterizes a one-parameter family of self-adjoint extensions of the
operator $L$ to the domain of differentiable functions obeying the boundary 
condition
\begin{equation}
\Psi(r,2 \pi) = \exp(i \theta) \Psi(r,0). 
\end{equation}
Since the coordinates $\varphi = 0$ and $\varphi = 2 \pi$ describe the same 
physical point on the cone, the requirement of single-valuedness of the 
physical wave function restricts us to $\theta = 0$. Hence, for wave functions 
on the cone the domain ${\cal D}[L] \in {\cal H}$ consists of the periodic
differentiable functions $\Psi$ (with $L \Psi \in {\cal H}$) which obey
\begin{equation}
\label{period}
\Psi(r,2 \pi) = \Psi(r,0).
\end{equation}

The operator for the kinetic energy takes the form
\begin{equation}
T = - \frac{1}{2M} \left(\p_r^2 + \frac{1}{r} \p_r + 
\frac{1}{r^2 s^2} \p_\varphi^2\right).
\end{equation}
Since $\p_\varphi = s \p_\chi$, this operator seems to be identically the same 
as the standard one operating on wave functions on the plane. However, in order
to completely define $T$, one must again identify the domain ${\cal D}[T]$ of 
wave functions on which it acts. First of all, these functions should again 
obey eq.(\ref{period}), i.e.\ they must be periodic in the rescaled angle 
$\varphi$ (not in the original polar angle $\chi$ of the full plane). 
Separating the angular dependence
\begin{equation}
\Psi(r,\varphi) = \psi(r) \exp(i m \varphi),
\end{equation}
eq.(\ref{period}) leads to $m \in \Z$ as well as to the kinetic energy
\begin{equation}
T = - \frac{1}{2M} \left(\p_r^2 + \frac{1}{r} \p_r\right) + 
\frac{m^2}{2 M r^2 s^2}.
\end{equation}
Effectively, a positive deficit angle $\delta$ (i.e.\ $s < 1$) leads to an
enhancement of the centrifugal barrier, while a negative deficit angle
($s > 1$) leads to its reduction.

Let us now address the issues of Hermiticity and self-adjointness of $T$. First
of all, the radial wave functions belong to the radial Hilbert space 
${\cal H}_r = L_2((0,\infty);r)$. It is well-known that the operator 
$- i \p_r$ is not even Hermitean in ${\cal H}_r$. Indeed, the 
Hermitean conjugate of $\p_r$ is
\begin{equation}
\label{drdagger}
\p_r^\dagger = - \p_r - \frac{1}{r}.
\end{equation}
This follows from 
\begin{eqnarray}
\langle \phi|\p_r \psi\rangle&=&
\int_0^\infty dr \ r \phi(r)^* \p_r \psi(r) \nonumber \\
&=&- \int_0^\infty dr \ \p_r \left[r \phi(r)^*\right] \psi(r) +
r \phi(r)^* \psi(r)|_0^\infty \nonumber \\
&=&- \int_0^\infty dr \ \left[r \p_r \phi(r)^* + \phi(r)^*\right] \psi(r) +
r \phi(r)^* \psi(r)|_0^\infty \nonumber \\
&=&\int_0^\infty dr \ r 
\left[ - \p_r \phi(r)^* - \frac{1}{r} \phi(r)^*\right] \psi(r) +
r \phi(r)^* \psi(r)|_0^\infty \nonumber \\
&=&\langle \p_r^\dagger \phi|\psi\rangle + r \phi(r)^* \psi(r)|_0^\infty.
\end{eqnarray}
Since the partial integration should not lead to boundary terms, we must 
require $r \phi(r)^* \psi(r)|_0^\infty = 0$. It is natural to define the 
operator 
\begin{equation}
D_r = - i \left(\p_r + \frac{1}{2 r}\right) = 
- i \frac{1}{\sqrt{r}} \p_r \sqrt{r}.
\end{equation}
In the domain ${\cal D}[D_r]$ of differentiable functions $\psi(r)$ (with 
$D_r \psi \in {\cal H}_r$) obeying $\psi(0) = 0$, the operator $D_r$ is indeed 
Hermitean because formally
\begin{equation}
D_r^\dagger = i \left(\p_r^\dagger + \frac{1}{2 r}\right) = 
i \left(- \p_r  - \frac{1}{r} + \frac{1}{2 r}\right) = 
- i \left(\p_r + \frac{1}{2 r}\right).
\end{equation}
However, it does not represent a proper physical observable because it is not 
self-adjoint. It is interesting to note that
\begin{equation}
D_r^2 = - \left(\p_r + \frac{1}{2r} \right)^2 = - \p_r^2 - \frac{1}{r} \p_r +
\frac{1}{4r^2},
\end{equation}
which is closely related to the kinetic energy operator $T$, possesses a 
family of self-adjoint extensions. Eq.(\ref{drdagger}) also seems to readily
imply Hermiticity of the kinetic energy operator $T$ because, at least 
formally,
\begin{eqnarray}
\left(\p_r^2 + \frac{1}{r} \p_r\right)^\dagger&=& 
\p_r^{\dagger 2} + \p_r^\dagger \frac{1}{r} =
\left(\p_r + \frac{1}{r}\right)^2 - 
\left(\p_r + \frac{1}{r}\right) \frac{1}{r} \nonumber \\
&=&\p_r^2 + \frac{2}{r} \p_r - \frac{1}{r} \p_r = \p_r^2 + \frac{1}{r} \p_r.
\end{eqnarray}
However, the issue is again more subtle because one should consider
\begin{eqnarray}
\langle \phi|\left(\p_r^2 + \frac{1}{r} \p_r\right) \psi\rangle&=&
\int_0^\infty dr \ r \phi(r)^* \left(\p_r^2 + \frac{1}{r} \p_r\right) \psi(r)
\nonumber \\
&=&- \int_0^\infty dr \ \p_r \left[r \phi(r)^*\right] \p_r \psi(r) +
r \phi(r)^* \p_r \psi(r)|_0^\infty \nonumber \\
&-&\int_0^\infty dr \ \p_r \phi(r)^* \psi(r) + \phi(r)^* \psi(r)|_0^\infty 
\nonumber \\
&=&\int_0^\infty dr \ \p_r^2 \left[r \phi(r)^*\right] \psi(r) -
\p_r \left[r \phi(r)^*\right] \psi(r)|_0^\infty \nonumber \\
&-&\int_0^\infty dr \ \p_r \phi(r)^* \psi(r) + 
\left[r \phi(r)^* \p_r \psi(r) + \phi(r)^* \psi(r)\right]_0^\infty 
\nonumber \\
&=&\int_0^\infty dr \ r 
\left[\left(\p_r^2 + \frac{2}{r} \p_r - \frac{1}{r} \p_r\right)
\phi(r)^*\right] \psi(r) \nonumber \\
&+&\left[r \phi(r)^* \p_r \psi(r) - r \p_r \phi(r)^* \psi(r)\right]_0^\infty
\nonumber \\
&=&\langle \left(\p_r^2 + \frac{1}{r} \p_r \right) \phi|\psi\rangle +
\left[r \phi(r)^* \p_r \psi(r) - r \p_r \phi(r)^* \psi(r)\right]_0^\infty.
\nonumber \\ \,
\end{eqnarray}
Hence, in order to ensure the Hermiticity of $T$, one must impose the condition
\begin{equation}
\label{cond}
\left[r \phi(r)^* \p_r \psi(r) - r \p_r \phi(r)^* \psi(r)\right]_0^\infty = 0, 
\end{equation}
which admits a one-parameter family of self-adjoint extensions. The 
self-adjoint extensions of $T$ have been studied in \cite{Kay91}. It turns out 
that the tip of the cone is a singular point that may be endowed with 
non-trivial physical properties. These properties are described by a 
real-valued parameter that defines a family of self-adjoint extensions. 
Physically speaking, the different self-adjoint extensions correspond to 
properly renormalized $\delta$-function potentials of different strengths 
located at the tip of the cone. In this paper, we limit ourselves to the case 
without $\delta$-function potentials, which corresponds to the so-called 
Friedrichs extension \cite{Ree72} characterized by the boundary condition
\begin{equation}
\label{boundary}
\lim_{r \rightarrow 0} r \p_r \psi(r) = 0.
\end{equation}
If we impose this condition on $\psi \in {\cal D}[T]$ and also want to satisfy 
eq.(\ref{cond}), the function $\phi \in {\cal D}[T^\dagger]$ must also obey 
eq.(\ref{boundary}). As a result, ${\cal D}[T^\dagger] = {\cal D}[T]$, such 
that $T = T^\dagger$ is indeed self-adjoint.

While the cone is as flat as the plane, its singular tip and its deficit angle 
$\delta$ have drastic effects on the dynamics. In the following, we will 
consider a particle moving on a cone and bound to its tip by a $1/r$ or $r^2$ 
potential. Interestingly, when the deficit angle divided by $2 \pi$ (or 
equivalently $s$) is a rational number, all bound classical orbits are again
closed and once more there are additional degeneracies in the discrete spectrum
of the Hamilton operator $H$. Just like in the plane, the $1/r$ and $r^2$ 
potentials on a cone have accidental $SU(2)$ symmetries. However, unlike in the
plane, the corresponding multiplets may now have fractional ``spin'' and 
unusual degeneracies. This unusual behavior arises because, in this case, the
Runge-Lenz vector $\vec R$ --- although Hermitean in its appropriate domain 
${\cal D}[\vec R]$ ---  does not act as a Hermitean operator in the domain 
${\cal D}[H]$ of the Hamiltonian and thus does not represent a proper physical 
observable.

It should be noted that motion on a cone may not be an entirely academic 
problem. First, space-times with a conical singularity arise in the study of
cosmic strings. Indeed, $1/r$ and $r^2$ potentials have already been considered
in this context \cite{Fur00,Mar02}, however, without discussing accidental 
symmetries. Furthermore, graphene --- a single sheet of graphite, i.e.\ a 
honeycomb of carbon hexagons --- can be bent to form cones by adding or 
removing a wedge of carbon atoms and by replacing one hexagon by a carbon 
hepta- or pentagon \cite{Aze98,Sit07}. While the low-energy degrees of freedom 
in graphene are massless Dirac fermions, in this paper we limit ourselves to 
studying the Schr\"odinger equation for motion on a cone.

The rest of the paper is organized as follows. In section 2 we discuss the 
$1/r$ and in section 3 we discuss the $r^2$ potential on a cone. In both cases,
we construct the Runge-Lenz vector which, together with the angular momentum, 
generates the accidental $SU(2)$ symmetry. We also relate the Hamilton operator
to the corresponding Casimir operator and we discuss the unusual multiplets 
realized in the discrete spectrum. Section 4 contains our conclusions.

\section{The $1/r$ Potential on a Cone}

In this section we consider a particle on the surface of a cone bound to its 
tip by a $1/r$ potential
\begin{equation}
V(r) = - \frac{\kappa}{r}.
\end{equation}
The corresponding total energy is thus given by
\begin{equation}
H = T + V =
\frac{1}{2M} \left(p_r^2 + \frac{L^2}{r^2 s^2}\right) - \frac{\kappa}{r}.
\end{equation}

\subsection{Classical Solutions}

It is straightforward to find the most general bound solution of the classical
equations of motion and one obtains the classical orbit
\begin{equation}
\label{orbitr}
\frac{1}{r} = \frac{M \kappa s^2}{L^2} 
\left[1 + e \cos(s(\varphi - \varphi_0)) \right],
\end{equation}
with the eccentricity given by
\begin{equation}
\label{eccentricity}
e = \sqrt{1 + \frac{2 E L^2}{M \kappa^2 s^2}},
\end{equation}
where $E < 0$ is the energy and $L$ is the angular momentum. The radial 
component of the momentum takes the form
\begin{equation}
\label{orbitp}
p_r = \frac{M \kappa s}{L} e \sin(s(\varphi - \varphi_0)).
\end{equation}
The angle $\varphi_0$ determines the direction of the perihelion. Obviously, 
the classical orbit is closed as long as $s = p/q$ is a rational number (with
$p, q \in \N$ not sharing a common divisor). In that case, after $q$ 
revolutions around the tip of the cone, both $r$ and $p_r$ return to their 
initial values. Some examples of classical orbits are shown in figure 2.
\begin{figure}[htb]
\begin{center}
\epsfig{file=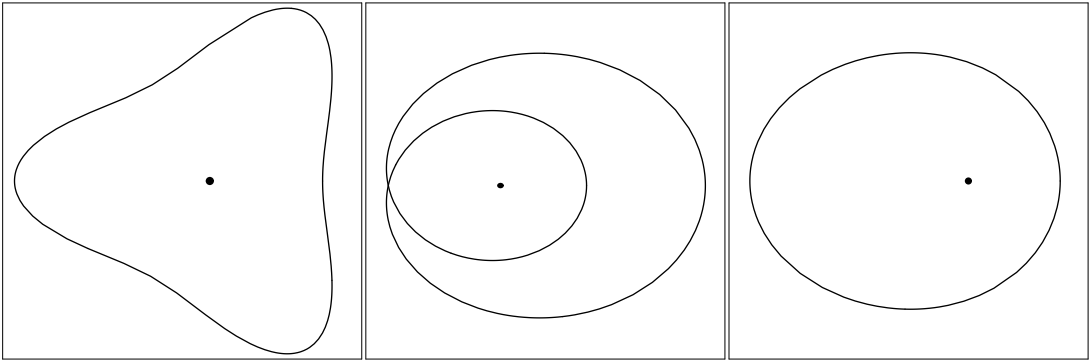,width=15cm}
\end{center}
\caption{\it Examples of bound classical orbits for the $1/r$ potential with
$s = 3$ (left), $s = \frac{1}{2}$ (middle), and $s = 1$ (right). The latter 
case represents a standard Kepler ellipse. The orbits are shown in the 
$x$-$y$-plane with $(x,y) = r (\cos\varphi,\sin\varphi)$ where 
$\varphi = \chi/s \in [0,2 \pi]$ is the rescaled polar angle.}
\end{figure}

\subsection{Semi-classical Bohr-Sommerfeld Quantization}

Let us consider Bohr-Sommerfeld quantization. The quantization condition for 
the angular momentum takes the form
\begin{equation}
\oint d\varphi \ L = 2 \pi L = 2 \pi m,
\end{equation}
such that $L = m \in \Z$. Similarly, the quantization condition for the radial 
motion is given by
\begin{equation}
\oint dr \ p_r = 2 \pi \left(n_r + \frac{1}{2}\right), \ n_r \in \{0,1,2,...\}.
\end{equation}
The factor $1/2$, which is sometimes not taken into account in Bohr-Sommerfeld
quantization, arises for librations but is absent for rotations. Using 
eqs.(\ref{orbitr}), (\ref{eccentricity}), and (\ref{orbitp}) and integrating 
over the period $2 \pi/s$ it is straightforward to obtain
\begin{equation}
\oint dr \ p_r = \int_0^{2 \pi/s} d\varphi \ 
\frac{|L| e^2 \sin^2(s(\varphi - \varphi_0))}
{\left(1 + e \cos(s(\varphi - \varphi_0)\right)^2} = 
2 \pi \left(\sqrt{- \frac{M \kappa^2}{2 E}} - \frac{|L|}{s}\right),
\end{equation}
which leads to
\begin{equation}
\label{energyC}
E = - \frac{M \kappa^2}{2 \left(n_r + \frac{|m|}{s} + \frac{1}{2}\right)^2}.
\end{equation}
It will turn out that this result is exact and not just limited to the 
semi-classical regime. 

\subsection{Solution of the Schr\"odinger Equation}

The radial Schr\"odinger equation takes the form
\begin{equation}
\left[- \frac{1}{2M} \left(\p_r^2 + \frac{1}{r} \p_r\right) + 
\frac{m^2}{2 M r^2 s^2} - \frac{\kappa}{r}\right] \psi(r) = E \psi(r),
\end{equation}
which is straightforward to solve. Using $\alpha = \sqrt{8 M |E|}$, one finds
\cite{Fur00}
\begin{equation}
\label{wavefunction}
\psi_{n_r,m}(r) = A \exp(- \frac{\alpha r}{2}) (\alpha r)^{|m|/s} 
{_1 F_1}(- n_r,\frac{2 |m|}{s} + 1,\alpha r),
\end{equation}
where ${_1 F_1}$ is a confluent hyper-geometric function. The corresponding 
quantized energy values are given by eq.(\ref{energyC}). Now $n_r$ is the 
number of nodes of the radial wave function and $m \in \Z$ is the angular 
momentum quantum number. Parity symmetry together with the $SO(2)$ rotational 
symmetry ensures the degeneracy of states with quantum numbers $m$  and $- m$. 
It is interesting to note that there are additional accidental degeneracies in 
the spectrum when the scale factor $s$ is a rational number. Remarkably, this 
is just the condition under which all classical orbits are closed. It is worth
mentioning that there are other mathematical solutions of the Schr\"odinger
equations which, however, do not qualify as physical wave functions because 
they diverge at the origin. These solutions will play a certain role later, 
when we discuss the accidental symmetry multiplets.

For example, let us consider the case $s = \frac{1}{2}$ corresponding to the
deficit angle $\delta = \pi$. In that case, a wave function without nodes 
(i.e.\ with $n_r = 0$) and with angular momentum $m = \pm 1$ is 
degenerate in energy with a wave function with two nodes ($n_r = 2$) and with 
$m = 0$. As another example, let us consider $s = 2$ which corresponds to the 
negative deficit angle $\delta = - 2 \pi$. In this case, one builds a ``cone'' 
by cutting two planes open and gluing them together in the same way as the 
double-layered Riemann surface of the complex square root. This effectively 
lowers the centrifugal barrier by a factor of $s^2 = 4$. In this case, a wave 
function without nodes ($n_r = 0$) and with $m = \pm 2$ is degenerate with a 
wave function with one node ($n_r = 1$) and with $m = 0$. Similarly, for 
$s = n \in \N$, one glues $n$ cut planes to a ``cone'' in the same way as the 
multi-layered Riemann surface of the complex $n$-th root. Now, a wave function 
without nodes ($n_r = 0$) and with $m = \pm n$ is degenerate with a wave 
function with one node ($n_r = 1$) and with $m = 0$. Some features of the 
energy spectrum are illustrated in figure 3.
\begin{figure}[htb]
\begin{center}
\epsfig{file=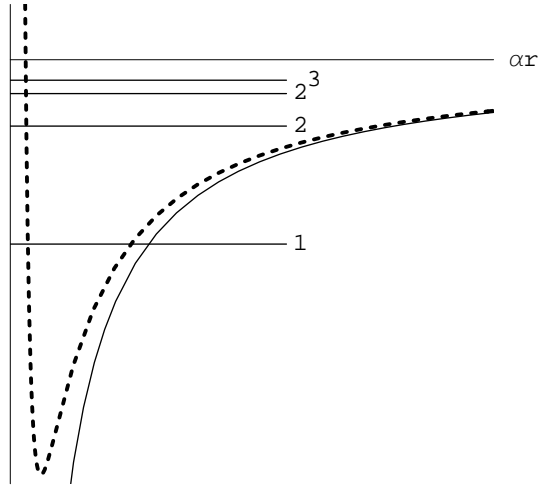,width=7cm}
\end{center}
\caption{\it The $1/r$ potential (solid curve) together with an effective
potential including the centrifugal barrier with $m = \pm 1$ (dashed curve) for
$s = 3$. The energies of the ground state and the first three excited states 
are indicated by horizontal lines. The numbers besides the lines specify the
degree of degeneracy. The ground state (with $n_r = 0, m = 0$) is
non-degenerate, while the first and second excited states (with 
$n_r = 0, m = \pm 1$ and $n_r = 0, m = \pm 2$, respectively) are two-fold 
degenerate due to parity symmetry. The third excited level has an accidental 
three-fold degeneracy and consists of the states with $n_r = 0, m = \pm 3$ and 
$n_r = 1, m = 0$.}
\end{figure}

\subsection{Runge-Lenz Vector and $SU(2)$ Algebra}

The fact that, for rational $s$, all classical orbits are closed and the 
discrete spectrum of the Hamiltonian has accidental degeneracies suggests that 
there is a hidden conserved quantity generating a corresponding accidental 
symmetry. From the plane (with $s = 1$) we are familiar with the Runge-Lenz 
vector and an accidental $SU(2)$ symmetry. Indeed a Runge-Lenz vector can be 
constructed for any $s$. However, it should be pointed out that, due to the 
conical geometry, the resulting object no longer transforms as a proper vector.
We still continue to refer to it as the ``Runge-Lenz vector''.

At the classical level, we can use eqs.(\ref{orbitr}) and (\ref{orbitp}) to 
write
\begin{eqnarray}
&&\kappa e \cos(s(\varphi - \varphi_0)) = 
\kappa e \left[\cos(s \varphi) \cos(s \varphi_0) + 
\sin(s \varphi) \sin(s \varphi_0)\right] = \frac{L^2}{M r s^2} - \kappa, 
\nonumber \\
&&\kappa e \sin(s(\varphi - \varphi_0)) = 
\kappa e \left[\sin(s \varphi) \cos(s \varphi_0) -
\cos(s \varphi) \sin(s \varphi_0)\right] = \frac{p_r L}{M s},
\end{eqnarray}
such that
\begin{eqnarray}
&&R_x = \kappa e \cos(s \varphi_0) = 
\left(\frac{L^2}{M r s^2} - \kappa\right) \cos(s \varphi) + 
\frac{p_r L}{M s} \sin(s \varphi), \nonumber \\
&&R_y = \kappa e \sin(s \varphi_0) = 
\left(\frac{L^2}{M r s^2} - \kappa\right) \sin(s \varphi) - 
\frac{p_r L}{M s} \cos(s \varphi),
\end{eqnarray}
are indeed independent of time. Furthermore, for $s = 1$, $R_x$ and $R_y$ are
just the components of the familiar Runge-Lenz vector. It should be noted that,
for non-integer values of $s$, the quantities $R_x$ and $R_y$ are not conserved
quantities in the usual sense. In particular, they are not single-valued 
functions of the coordinates $x = r \cos\varphi$ and $y = r \sin\varphi$, but
depend on the angle $\varphi$ itself. As a consequence, the values of $R_x$ 
and $R_y$ depend on the history of the motion, i.e.\ on the number of 
revolutions around the tip of the cone. However, quantities that are 
``conserved'' only because they refer back to the initial conditions, do not 
qualify as proper physical constants of motion. To further clarify this issue, 
it is useful to construct the complex variables
\begin{equation}
R_\pm = R_x \pm i R_y = \left(\frac{L^2}{M r s^2} - \kappa
\mp i \frac{p_r L}{M s}\right) \exp(\pm i s \varphi).
\end{equation}
For rational values $s = p/q$ (with $p, q \in \N$) the quantities
\begin{equation}
R_\pm^q = \left(\frac{L^2}{M r s^2} - \kappa
\mp i \frac{p_r L}{M s}\right)^q \exp(\pm i p \varphi)
\end{equation}
are single-valued functions of $x = r \cos\varphi$ and $y = r \sin\varphi$, and
hence qualify as proper conserved quantities. Remarkably, exactly for rational 
values of $s$ all bound classical orbits are closed.

The length of the Runge-Lenz vector is given by
\begin{eqnarray}
R^2&=&R_x^2 + R_y^2 = \left(\frac{L^2}{M r s^2} - \kappa\right)^2 + 
\left(\frac{p_r L}{M s}\right)^2 \nonumber \\
&=&2 \frac{L^2}{M s^2} \left(\frac{p_r^2}{2 M} + \frac{L^2}{2 M r^2 s^2} 
- \frac{\kappa}{r}\right) + \kappa^2 = 
\frac{2 H L^2}{M s^2} + \kappa^2.
\end{eqnarray}
In the quantum mechanical treatment it will turn out to be useful to introduce
the rescaled variables
\begin{equation}
\label{rescaling}
\widetilde R_x = \sqrt{- \frac{M}{2 H}} R_x, \
\widetilde R_y = \sqrt{- \frac{M}{2 H}} R_y, \ \widetilde L = \frac{L}{s},
\end{equation}
which makes sense for bound orbits with negative energy. We then obtain
\begin{equation}
C = \widetilde R_x^2 + \widetilde R_y^2 + \widetilde L^2 = - 
\frac{M \kappa^2}{2 H} \ \Rightarrow \ H = - \frac{M \kappa^2}{2 C}.
\end{equation}
It will turn out that the quantum analogue of $C$ is the Casimir operator of an
accidental $SU(2)$ symmetry.

At the quantum level, the components of the Runge-Lenz vector turn into
operators
\begin{eqnarray}
\label{eqRpm}
R_x&=&- \frac{1}{M r s^2} \cos(s \varphi) \p_\varphi^2 +
\frac{1}{2 M r s} \sin(s \varphi) \p_\varphi - \kappa \cos(s \varphi) 
\nonumber \\
&-&\frac{1}{M s} \sin(s \varphi) \p_r \p_\varphi - 
\frac{1}{2 M} \cos(s \varphi) \p_r, \nonumber \\
R_y&=&- \frac{1}{M r s^2} \sin(s \varphi) \p_\varphi^2 -
\frac{1}{2 M r s} \cos(s \varphi) \p_\varphi - \kappa \sin(s \varphi) 
\nonumber \\
&+&\frac{1}{M s} \cos(s \varphi) \p_r \p_\varphi -
\frac{1}{2 M} \sin(s \varphi) \p_r.
\end{eqnarray}
It is straightforward (but somewhat tedious) to show that not only the angular
momentum $L$, but also both components of the Runge-Lenz vector commute with 
the Hamiltonian, i.e.
\begin{equation}
[R_x,H] = [R_y,H] = [L,H] = 0,
\end{equation}
and that these operators obey the algebra
\begin{equation}
[R_x,R_y] = - i \frac{2 HL}{M s}, \ [R_x,L] = - i s R_y, \ 
[R_y,L] = i s R_x.
\end{equation}
Applying the rescaling of eq.(\ref{rescaling}), this leads to
\begin{equation}
\label{SU2}
[\widetilde R_x,\widetilde R_y] = i \widetilde L, \
[\widetilde R_y,\widetilde L] = i \widetilde R_x, \
[\widetilde L,\widetilde R_x] = i \widetilde R_y.
\end{equation}
Hence, $\widetilde R_x$, $\widetilde R_y$, and $\widetilde L$ generate an 
$SU(2)$ algebra.

\subsection{Casimir Operator}

It is straightforward to construct the Casimir operator of the $SU(2)$ algebra
and one obtains
\begin{equation}
C = \widetilde R_x^2 + \widetilde R_y^2 + \widetilde L^2 = 
- \frac{M \kappa^2}{2 H} - \frac{1}{4},
\end{equation}
such that 
\begin{equation}
H = - \frac{M \kappa^2}{2 \left(C + \frac{1}{4}\right)} =  
- \frac{M \kappa^2}{2 \left(S + \frac{1}{2}\right)^2}.
\end{equation}
In the last step we have used the fact that the eigenvalue of the Casimir 
operator of an $SU(2)$ symmetry is naturally represented by $S(S + 1)$ such 
that
\begin{equation}
C + \frac{1}{4} = S(S + 1) + \frac{1}{4} = \left(S + \frac{1}{2}\right)^2.
\end{equation}
By comparison with eq.(\ref{energyC}) for the energy spectrum, we thus identify
\begin{equation}
S = n_r + \frac{|m|}{s}.
\end{equation}
This result is puzzling, because for $2 |m|/s \notin \N$ the abstract spin $S$ 
is not an integer or a half-integer. For a general scale factor $s$ (or 
equivalently for general deficit angle $\delta$), the abstract spin is, in 
fact, continuous. Even for general rational $s$, for which all bound classical 
orbits are closed and there are accidental degeneracies in the discrete 
spectrum of the Hamiltonian, the spin $S$ is not just an integer or a 
half-integer.

\subsection{Domains of Operators and Hermiticity}

In order to better understand the puzzling result that the Casimir ``spin'' may
not be an integer or half-integer, let us address the questions of Hermiticity 
and of the domains of the various operators. As we have discussed in the 
introduction, once it is endowed with an appropriate extension, the Hermitean 
kinetic energy operator $T$ becomes self-adjoint and thus qualifies as a 
physical observable. The same is true for the full Hamiltonian including the 
potential. In this case, we assume the standard Friedrichs extension
\cite{Ree72}, which implies that there is no $\delta$-function potential 
located at the tip of the cone.

Using $\p_r^\dagger = - \p_r - 1/r$ as well as 
$\p_\varphi^\dagger = - \p_\varphi$, it is straightforward to show that, at 
least formally, $\widetilde R_x^\dagger = \widetilde R_x$ and 
$\widetilde R_y^\dagger = \widetilde R_y$, which implies
$\widetilde R_\pm^\dagger = \widetilde R_\mp$. However, as we have seen in the
introduction, Hermiticity also requires appropriate boundary conditions, which
restrict the domains of the corresponding operators. It is interesting to note
that, using $s \varphi = \chi$, the operators $R_x$ and $R_y$ of 
eq.(\ref{eqRpm}) formally agree with the components of the standard Runge-Lenz 
vector for the plane from eq.(\ref{RungeLenz}). The Runge-Lenz vector for the
plane is a Hermitean and even self-adjoint operator acting in a domain
${\cal D}[\vec R]$ that contains the domain of the Hamiltonian. This domain 
contains smooth functions which are $2 \pi$-periodic in the polar angle $\chi$ 
of the plane. The operators $\widetilde R_x$ and $\widetilde R_y$, on the other
hand, act on the Hilbert space of square-integrable wave functions on the cone.
In this case, the domain of the Hamiltonian ${\cal D}[H]$ contains smooth 
functions which are $2 \pi$-periodic in the rescaled angle $\varphi$ and obey 
the boundary condition of eq.(\ref{boundary}). While $\widetilde R_x$ and 
$\widetilde R_y$ on the cone are still Hermitean in their appropriate domain, 
in contrast to the case of the plane, they are not Hermitean in the domain 
${\cal D}[H]$ of the Hamiltonian. In particular, for $s \neq 1$ the operators 
$\widetilde R_x$ and $\widetilde R_y$ map $2 \pi$-periodic physical wave 
functions onto functions outside ${\cal D}[H]$, because they contain 
multiplications with the $2 \pi/s$-periodic functions $\cos(s \varphi)$ and 
$\sin(s \varphi)$. Proper symmetry generators should map wave functions from 
the domain of the Hamiltonian back into ${\cal D}[H]$. Hence, for 
$s \notin \N$, the operators $\widetilde R_x$ and $\widetilde R_y$ do not 
represent proper symmetry generators. 

It is interesting to consider the case of rational $s = p/q$ with 
$p, q \in \N$. In this case, a single application of 
\begin{equation}
\widetilde R_\pm = \widetilde R_x \pm i \widetilde R_y
\end{equation}
may take us out of the domain of the Hamiltonian, but a $q$-fold application of
these operators brings us back into ${\cal D}[H]$. Indeed, just as for 
rational $s$ the classical object $R_\pm^q$ represents a proper physical
conserved quantity, $\widetilde R_\pm^q$ (but not $\widetilde R_\pm$ itself) 
qualifies as a proper symmetry generator. The case of integer $s = n$ is also 
interesting, because in that case $\cos(s \varphi)$ and $\sin(s \varphi)$ are 
indeed $2 \pi$-periodic. Hence, by acting with $\widetilde R_\pm$ we might
expect to stay within ${\cal D}[H]$, although for $n \geq 3$ the abstract spin 
$S = n_r + |m|/s = n_r + |m|/n$ is still quantized in unusual fractional units.
However, as we will see below, another subtlety arises because 
$\widetilde R_\pm$ may turn a physical wave function that is regular at the
origin (and thus obeys the boundary condition of eq.(\ref{boundary})) into a 
singular one. This further limits the domain of the operators
$\widetilde R_\pm$. The unusual (not properly quantized) value of the Casimir 
spin can be traced back to the mathematical fact that the Runge-Lenz vector ---
although Hermitean in its appropriate domain --- does not act as a Hermitean 
operator in the domain of the Hamiltonian. Hence, in retrospect the $SU(2)$ 
commutation relations of eq.(\ref{SU2}) are rather formal. In fact, they are 
satisfied for functions $\Psi(r,\varphi)$ with $\varphi \in \R$, but not for 
the periodic functions in ${\cal D}[H]$ for which $\varphi \in [0,2 \pi]$. This
is another indication that the accidental ``$SU(2)$'' symmetry of 
eq.(\ref{SU2}) is rather unusual.

\subsection{Unusual Multiplets}

How can we further understand the puzzling result that the Casimir spin $S$ is 
not always quantized in integer or half-integer units? It seems that we found 
new unusual representations for something as well understood as an $SU(2)$ 
algebra. Acting with $\widetilde R_\pm$ on a $2 \pi$-periodic wave function 
\begin{equation}
\bra r,\varphi|n_r,m \ket = \psi_{n_r,m}(r) \exp(i m \varphi)
\end{equation}
one changes both $n_r \in \N$ and $m \in \Z$. For $m > 0$ one obtains
\begin{equation}
\label{eq1}
\widetilde R_+ |n_r,m \ket \propto |n_r - 1,m + s \ket, \
\widetilde R_- |n_r,m \ket \propto |n_r + 1,m - s \ket,
\end{equation}
and for $m < 0$ one finds
\begin{equation}
\label{eq2}
\widetilde R_+ |n_r,m \ket \propto |n_r + 1,m + s \ket, \
\widetilde R_- |n_r,m \ket \propto |n_r - 1,m - s \ket.
\end{equation}
Finally, for $m = 0$ we have
\begin{equation}
\label{eq3}
\widetilde R_+ |n_r,0 \ket \propto |n_r - 1,s \ket, \
\widetilde R_- |n_r,0 \ket \propto |n_r - 1,- s \ket.
\end{equation}
These relations follow from the $SU(2)$ algebra which implies that
$\widetilde R_\pm$ are raising and lowering operators for $\widetilde L = L/s$.
Hence, by acting with $\widetilde R_\pm$ the eigenvalue $m$ of $L$ is shifted
by $\pm s$. Using the fact that the eigenvalue of the Casimir operator, which 
is determined by $S = n_r + |m|/s$, does not change under applications of 
$\widetilde R_\pm$, one immediately obtains the effects of $\widetilde R_\pm$
on the radial quantum number $n_r$. Eqs.(\ref{eq1}), (\ref{eq2}), and
(\ref{eq3}) also follow directly by applying the explicit forms of 
$R_\pm = R_x \pm i R_y$ from eq.(\ref{eqRpm}) to the wave functions of 
eq.(\ref{wavefunction}).

Again, by using the wave functions of eq.(\ref{wavefunction}) one can show that
\begin{eqnarray}
&&\widetilde R_+^{n_r+1} |n_r,m \geq 0 \ket \propto 
\widetilde R_+ |0,m + n_r s \geq 0\ket = 0, \nonumber \\
&&\widetilde R_-^{n_r+1} |n_r,m \leq 0 \ket \propto 
\widetilde R_- |0,m - n_r s \leq 0\ket = 0.
\end{eqnarray}
Hence, depending on the sign of $m$, by acting $n_r + 1$ times either with 
$\widetilde R_+$ or with $\widetilde R_-$ we reach zero, and thus the multiplet
naturally terminates. This allows us to confirm the value of the Casimir spin
$S = n_r + |m|/s$ by evaluating
\begin{eqnarray}
C |0,m + n_r s \geq 0 \ket&=&\left[
\frac{1}{2} (\widetilde R_+ \widetilde R_- + \widetilde R_- \widetilde R_+) +
\widetilde L^2\right] |0,m + n_r s\geq 0 \ket \nonumber \\
&=&\left[
\frac{1}{2} ([\widetilde R_+,\widetilde R_-] + 2 \widetilde R_- \widetilde R_+)
+ \widetilde L^2\right] |0,m + n_r s\geq 0 \ket \nonumber \\
&=&(\widetilde L + \widetilde L^2) |0,m + n_r s\geq 0 \ket \nonumber \\
&=&\left(\frac{m}{s} + n_r\right)\left(\frac{m}{s} + n_r + 1\right) 
|0,m + n_r s\geq 0 \ket \nonumber \\
&=&S(S+1) |0,m + n_r s\geq 0 \ket, \nonumber \\
C |0,m - n_r s\leq 0 \ket&=&\left[
\frac{1}{2} (\widetilde R_+ \widetilde R_- + \widetilde R_- \widetilde R_+) +
\widetilde L^2\right] |0,m - n_r s\leq 0 \ket \nonumber \\
&=&\left[
\frac{1}{2} (\widetilde R_+ \widetilde R_- + 2 [\widetilde R_-,\widetilde R_-])
+ \widetilde L^2\right] |0,m - n_r s\leq 0 \ket \nonumber \\
&=&(- \widetilde L + \widetilde L^2) |0,m - n_r s \leq 0 \ket \nonumber \\
&=&\left(- \frac{m}{s} + n_r\right) \left(- \frac{m}{s} + n_r + 1\right) 
|0,m - n_r s\leq 0 \ket 
\nonumber \\
&=&S(S+1) |0,m - n_r s\leq 0 \ket.
\end{eqnarray}
The multiplet of degenerate states with the same value of $S$ can now be
obtained by $n$ repeated applications of either $\widetilde R_+$ or 
$\widetilde R_-$. It is important to note that, if $s$ is not an integer, 
$m \pm n s$ may also not be an integer and thus the corresponding state
may be outside ${\cal D}[H]$. Despite this, its radial wave function is still 
defined by eq.(\ref{wavefunction}) and it still solves the radial Schr\"odinger
equation. 

Let us first consider the generic case of irrational $s$. In that case, the 
classical orbits are not closed, there are no accidental degeneracies in the
discrete spectrum of the Hamiltonian, and the Casimir spin
$S = n_r + |m|/s$ is irrational. Acting with $\widetilde R_\pm$ on the 
$2 \pi$-periodic wave function $|n_r,m \in \Z \ket$ an arbitrary number of 
times, one generates functions which are not $2 \pi$-periodic and thus outside 
${\cal D}[H]$. As a consequence of parity symmetry, for $m \neq 0$ the two 
levels with the quantum numbers $m$ and $- m$ are still degenerate. However, 
that two-fold degeneracy is not accidental.

Next, let us discuss the case of rational $s = p/q$ in which all classical 
orbits are closed and there are accidental degeneracies in the discrete 
spectrum of the Hamiltonian. First, we consider the case 
$2 |m|/s = 2 |m| q/p \in \N$ for which the Casimir spin $S$ is an integer or a 
half-integer. Only in that case, the set of degenerate wave functions 
terminates on both ends, i.e.\
\begin{equation}
\widetilde R_-^{2 S + 1} |0,m + n_r s \geq 0 \ket = 0, \
\widetilde R_+^{2 S + 1} |0,m - n_r s \leq 0 \ket = 0.
\end{equation}
This follows by applying the operators of eq.(\ref{eqRpm}) to the wave 
function of eq.(\ref{wavefunction}), which is a somewhat tedious procedure 
that requires using non-trivial properties of the confluent hyper-geometric
functions. Our experience with $SU(2)$ algebras would 
suggest that there are $2 S + 1$ degenerate states. However, we should not 
forget that a single application of the raising and lowering operators
$\widetilde R_\pm$ may take us outside ${\cal D}[H]$, and only $q$ applications
of $\widetilde R_\pm$ take us back into ${\cal D}[H]$. It is straightforward 
(but not very illuminating) to count the number of states inside ${\cal D}[H]$.
In any case, for $s \neq 1$ this number is smaller than the naively expected 
$2 S + 1$. When $s S \in \N$, the degeneracy is given by $g = [2 S/q] + 1$, 
where $[2 S/q]$ denotes the nearest integer below $2 S/q$.

As we will see now, the multiplets are even more unusual in the case of 
rational $s = p/q$ with the Casimir spin $S$ neither being an integer nor a 
half-integer. In that case, the set of degenerate wave functions only 
terminates on one end, but not on the other. In particular, while still 
$\widetilde R_+|0,m + n_r s \geq 0 \ket = 0$, 
$\widetilde R_-^n|0,m + n_r s \geq 0 \ket$ does not 
vanish, even for arbitrarily large $n$.
Since an infinite number of values $m + (n_r - n) s$ will be integers, one 
might then think that the multiplet of degenerate states inside ${\cal D}[H]$ 
should contain an infinite number of states. Interestingly, this is not the 
case for a rather unusual reason. For $S$ neither being an integer nor a 
half-integer, the states $\widetilde R_-^n|0,m + n_r s \geq 0 \ket$ with 
$m + (n_r - n) s < 0$ are outside ${\cal D}[H]$ because the corresponding wave
function is singular at the origin. This again follows from applying the 
operators of eq.(\ref{eqRpm}) to the wave function of eq.(\ref{wavefunction}). 
Although they do not qualify as physical states, the divergent wave functions 
still are mathematical solutions of the Schr\"odinger differential equation 
which take the form
\begin{equation}
\psi(r) = A \exp(- \frac{\alpha r}{2}) (\alpha r)^{- |m|/s} 
{_1 F_1}(- n_r,- \frac{2 |m|}{s} + 1,\alpha r).
\end{equation}
The singularity of the wave function may or may not make the wave function 
non-normalizable. Even if it remains normalizable, the corresponding singular 
wave function does not belong to ${\cal D}[H]$ because it does not obey the 
boundary condition of eq.(\ref{boundary}). For $S$ neither being an integer nor
a half-integer, the states with positive and negative $m$ have the same energy
as a consequence of parity symmetry, but they are not related to one another by
applications of the raising and lowering operators $\widetilde R_\pm$.
Remarkably, in this case, by acting with a symmetry generator $\widetilde R_x$
or $\widetilde R_y$ on a wave function inside ${\cal D}[H]$, one may generate a
physically unacceptable wave function outside ${\cal D}[H]$. A sequence of
physical and unphysical wave functions is illustrated in figure 4.
\begin{figure}[htb]
\begin{center}
\epsfig{file=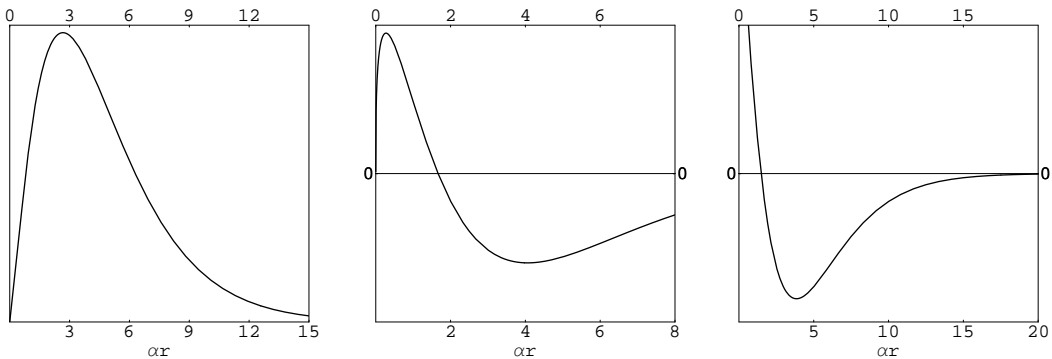,width=15cm}
\end{center}
\caption{\it A sequence of wave functions for the $1/r$ potential with $s = 3$
obtained from repeated applications of $\widetilde R_-$. The quantum numbers
are $n_r = 0$, $m = 4$ (left), $n_r = 1$, $m = 4 - s = 1$ (middle), and
$n_r = 2$, $m = 4 - 2s = -2$ (right). The third state in the sequence is 
outside the domain of the Hamiltonian because the corresponding wave function
does not obey the boundary condition of eq.(\ref{boundary}) and the state is 
thus unphysical.}
\end{figure}

To summarize, for $s \neq 1$ different types of unusual multiplets arise.
First, even for integer or half-integer $S = n_r + |m|/s$, the degeneracy of
the physical multiplet is not $2 S + 1$ because $m \pm n s$ may not be an 
integer in which case the corresponding wave function is not $2 \pi$-periodic. 
When $S = n_r + |m|/s$ is neither an integer nor a half-integer, there is an
infinite number of degenerate solutions of the Schr\"odinger equation. However,
only a finite number of them obeys the boundary condition of 
eq.(\ref{boundary}) and thus belongs to ${\cal D}[H]$.

\section{The $r^2$ Potential on a Cone}

Let us now turn to the problem of a particle moving on a cone and bound to its 
tip by a harmonic oscillator potential
\begin{equation}
V(r) = \frac{1}{2} M \omega^2 r^2.
\end{equation}
The Hamiltonian is then given by
\begin{equation}
H = T + V =
\frac{1}{2M} \left(p_r^2 + \frac{L^2}{r^2 s^2}\right)  +
\frac{1}{2} M \omega^2 r^2.
\end{equation}

\subsection{Classical Solutions}

Using the corresponding classical equations of motion one obtains the classical
orbits
\begin{equation}
\label{orbitrh}
\frac{1}{r^2}= \frac{M E s^2}{L^2}[1 + f \cos(2 s (\varphi - \varphi_0))],
\end{equation}
with $E$ and $L$ again denoting energy and angular momentum and with
\begin{equation}
f = \sqrt{1 - \frac{\omega^2 L^2}{E^2 s^2}}.
\end{equation}
The radial component of the momentum is given by
\begin{equation}
\label{orbitph}
\frac{p_r}{r} = \frac{M E s}{L} f \sin(2 s (\varphi - \varphi_0)).
\end{equation}
All classical orbits are closed as long as $2 s = p/q$  is a rational number
(with $p, q \in \N$ again not sharing a common divisor). Some examples of 
classical orbits are shown in figure 5.
\begin{figure}[htb]
\begin{center}
\epsfig{file=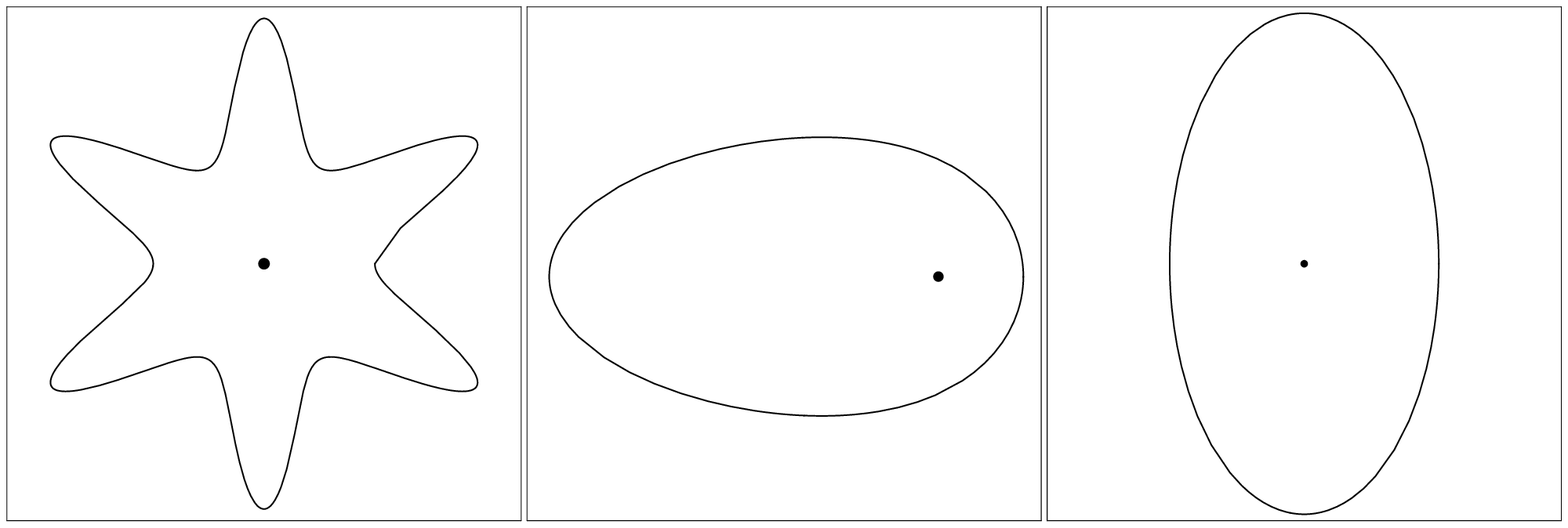,width=15cm}
\end{center}
\caption{\it Examples of bound classical orbits for the $r^2$ potential with
$s = 3$ (left), $s = \frac{1}{2}$ (middle), and $s = 1$ (right). The latter 
case represents an elliptic orbit of the standard harmonic oscillator. The 
orbits are shown in the $x$-$y$-plane with 
$(x,y) = r (\cos\varphi,\sin\varphi)$ where $\varphi = \chi/s \in [0,2 \pi]$ is
the rescaled polar angle.}
\end{figure}

\subsection{Semi-classical Bohr-Sommerfeld Quantization}

As in the case of the $1/r$ potential, the semi-classical quantization 
condition for the angular momentum is again given by $L = m \in \Z$. For the
harmonic oscillator the quantization condition for the radial motion takes the
form
\begin{equation}
\oint dr \ p_r = \int_0^{\pi/s} d\varphi \ 
\frac{|L| f^2 \sin^2(2 s(\varphi - \varphi_0))}
{\left(1 + f \cos(2 s(\varphi - \varphi_0)\right)^2} = 
\pi \left(\frac{E}{\omega} - \frac{|L|}{s}\right) = 2 \pi 
\left(n_r + \frac{1}{2}\right),
\end{equation}
such that
\begin{equation}
\label{energyH}
E = \omega \left(2 n_r + \frac{|m|}{s} + 1\right).
\end{equation}
Again, it will turn out that the semi-classical result exactly reproduces the 
one of the full quantum theory.

\subsection{Solution of the Schr\"odinger Equation}

For the particle on the cone with harmonic oscillator potential the radial
Schr\"odinger equation takes the form
\begin{equation}
\left[- \frac{1}{2M} \left(\p_r^2 + \frac{1}{r} \p_r\right) + 
\frac{m^2}{2 M r^2 s^2} + \frac{1}{2} M \omega^2 r^2\right] \psi(r) = 
E \psi(r).
\end{equation}
In this case, the solution is given by \cite{Fur00}
\begin{equation}
\psi_{n_r,m}(r) = A \exp(- \frac{\alpha^2 r^2}{2}) (\alpha r)^{|m|/s} 
{_1 F_1}(- n_r,\frac{|m|}{s} + 1,\alpha^2 r^2), \ \alpha = \sqrt{M \omega}.
\end{equation}
The corresponding quantized energy values are given by eq.(\ref{energyH}).
There are accidental degeneracies if $2 s = p/q$ is a rational number, which 
thus again arise exactly when all classical orbits are closed. Some features of
the energy spectrum are illustrated in figure 6.
\begin{figure}[htb]
\begin{center}
\epsfig{file=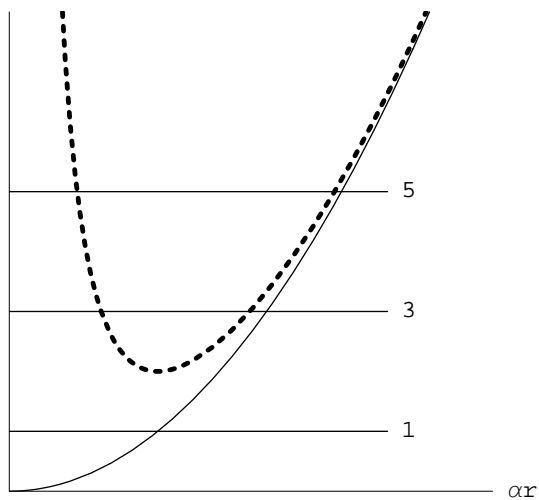,width=7cm}
\end{center}
\caption{\it The $r^2$ potential (solid curve) together with an effective
potential including the centrifugal barrier with $m = \pm 1$ (dashed curve) for
$s = \frac{1}{2}$. The energies of the ground state and the first two excited 
states are indicated by horizontal lines. The numbers besides the lines specify
the degree of degeneracy. The ground state (with $n_r = 0, m = 0$) is 
non-degenerate, while the first excited level (consisting of the states with 
$n_r = 0, m = \pm 1$ and $n_r = 1, m = 0$) and the second excited level
(consisting of the states with $n_r = 0, m = \pm 2$, $n_r = 1, m = \pm 1$, and
$n_r = 2, m = 0$) are accidentally three-fold, respectively five-fold 
degenerate.}
\end{figure}

\subsection{Runge-Lenz Vector}

The accidental degeneracies for rational $s$ again point to the existence of a
conserved Runge-Lenz vector. At the classical level, we can use 
eqs.(\ref{orbitrh}) and (\ref{orbitph}) to write
\begin{eqnarray}
&&\!\!\!\!\!\!\!\!\!\!E f \cos(2 s(\varphi - \varphi_0)) = 
E f \left[\cos(2 s \varphi) \cos(2 s \varphi_0) + 
\sin(2 s \varphi) \sin(2 s \varphi_0)\right] = \frac{L^2}{M r^2 s} - H
\nonumber \\
&&\!\!\!\!\!\!\!\!\!\!E f \sin(2 s(\varphi - \varphi_0)) = 
E f \left[\sin(2 s \varphi) \cos(2 s \varphi_0) -
\cos(2 s \varphi) \sin(2 s \varphi_0)\right] = \frac{p_r L}{M r s},
\end{eqnarray}
such that
\begin{eqnarray}
&&R_x = E f \cos(2 s \varphi_0) = 
\left(\frac{L^2}{M r^2 s^2} - H\right) \cos(2 s \varphi) + 
\frac{p_r L}{M r s} \sin(2 s \varphi), \nonumber \\
&&R_y = E f \sin(2 s \varphi_0) = 
\left(\frac{L^2}{M r^2 s^2} - H\right) \sin(2 s \varphi)
- \frac{p_r L}{M r s} \cos(2 s \varphi).
\end{eqnarray}
It should again be pointed out that $R_x$ and $R_y$ are proper conserved
quantities only if $2 s$ is an integer. Otherwise the Runge-Lenz vector is not
a $2 \pi$-periodic function of the angle $\varphi$, and its value depends on 
the number of revolutions of the particle around the tip of the cone. As 
before, it is useful to introduce the complex quantities 
\begin{equation}
R_\pm = R_x \pm i R_y = \left(\frac{L^2}{M r^2 s^2} - H
\mp i \frac{p_r L}{M r s}\right) \exp(\pm 2 i s \varphi).
\end{equation}
For rational values $2 s = p/q$ (with $p, q \in \N$) the quantities
\begin{equation}
R_\pm^q = \left(\frac{L^2}{M r^2 s^2} - H
\mp i \frac{p_r L}{M r s}\right)^q \exp(\pm i p \varphi)
\end{equation}
are again single-valued functions of $x = r \cos\varphi$ and 
$y = r \sin\varphi$, and are hence proper conserved quantities.

For the harmonic oscillator, the length of the Runge-Lenz vector is given by
\begin{eqnarray}
R^2&=&R_x^2 + R_y^2 = \left(\frac{L^2}{M r^2 s^2} - H\right)^2 + 
\left(\frac{p_r L}{M r s}\right)^2 \nonumber \\
&=&\left(\frac{p_r^2}{2 M} - \frac{L^2}{2 M r^2 s^2} + 
\frac{1}{2} M \omega^2 r^2 \right)^2 + \left(\frac{p_r L}{M r s}\right)^2 =
H^2 - \left(\omega \frac{L}{s}\right)^2.
\end{eqnarray}
As in the case of the $1/r$ potential, it is useful to introduce rescaled 
variables which now take the form
\begin{equation}
\label{rescalingh}
\widetilde R_x = \frac{1}{2 \omega} R_x, \
\widetilde R_y = \frac{1}{2 \omega} R_y, \ \widetilde L = \frac{L}{2 s}.
\end{equation}
We thus obtain
\begin{equation}
C = \widetilde R_x^2 + \widetilde R_y^2 + \widetilde L^2 = 
\left(\frac{H}{2 \omega}\right)^2 \ \Rightarrow \ H = 2 \omega \sqrt{C}.
\end{equation}
Once again, it will turn out that the quantum analogue of $C$ is the Casimir 
operator of an accidental $SU(2)$ symmetry.

At the quantum level the Runge-Lenz vector now takes the form
\begin{eqnarray}
R_x&=&\frac{1}{2M} \cos(2 s \varphi) \p_r^2
- \frac{1}{2 M r^2 s^2} \cos(2 s \varphi) \p_\varphi^2 +
\frac{1}{M r^2 s} \sin(2 s \varphi) \p_\varphi \nonumber \\
&-&\frac{1}{2} M \omega^2 r^2 \cos(2 s \varphi) -
\frac{1}{M r s} \sin(2 s \varphi) \p_r \p_\varphi - 
\frac{1}{2 M r} \cos(2 s \varphi) \p_r, \nonumber \\
R_y&=&\frac{1}{2M} \sin(2 s \varphi) \p_r^2
- \frac{1}{2 M r^2 s^2} \sin(2 s \varphi) \p_\varphi^2 -
\frac{1}{M r^2 s} \cos(2 s \varphi) \p_\varphi \nonumber \\
&-&\frac{1}{2} M \omega^2 r^2 \sin(2 s \varphi) +
\frac{1}{M r s} \cos(2 s \varphi) \p_r \p_\varphi - 
\frac{1}{2 M r} \sin(2 s \varphi) \p_r.
\end{eqnarray}
One can show that the Runge-Lenz vector as well as the angular momentum $L$ 
commute with the Hamiltonian, and that these operators obey the algebra
\begin{equation}
[R_x,R_y] = 2 i \omega \frac{L}{s}, \ [R_x,L] = - 2 i s R_y, \ 
[R_y,L] = 2 i s R_x.
\end{equation}
Applying the rescaling of eq.(\ref{rescalingh}), this leads to
\begin{equation}
[\widetilde R_x,\widetilde R_y] = i \widetilde L, \
[\widetilde R_y,\widetilde L] = i \widetilde R_x, \
[\widetilde L,\widetilde R_x] = i \widetilde R_y,
\end{equation}
which again represents an $SU(2)$ algebra.

\subsection{Casimir Operator}

The Casimir operator for the harmonic oscillator on the cone takes the form
\begin{equation}
C = \widetilde R_x^2 + \widetilde R_y^2 + \widetilde L^2 = 
\left(\frac{H}{2 \omega}\right)^2 - \frac{1}{4},
\end{equation}
which implies
\begin{equation}
H = 2 \omega \sqrt{C + \frac{1}{4}} = 2 \omega \left(S + \frac{1}{2}\right).
\end{equation}
Comparing with eq.(\ref{energyH}) for the energy spectrum, we now identify
\begin{equation}
S = n_r + \frac{|m|}{2 s}.
\end{equation}

\subsection{Unusual Multiplets}

Let us now consider the unusual multiplets in case of the harmonic oscillator.
The discussion is similar to the one of the $1/r$ potential and will thus not
be repeated in all details. For $m > 0$ one now obtains
\begin{equation}
\widetilde R_+ |n_r,m \ket \propto |n_r - 1,m + 2 s \ket, \
\widetilde R_- |n_r,m \ket \propto |n_r + 1,m - 2 s \ket,
\end{equation}
and for $m < 0$ one finds
\begin{equation}
\widetilde R_+ |n_r,m \ket \propto |n_r + 1,m + 2 s \ket, \
\widetilde R_- |n_r,m \ket \propto |n_r - 1,m - 2 s \ket,
\end{equation}
while, for $m = 0$ we have
\begin{equation}
\widetilde R_+ |n_r,0 \ket \propto |n_r - 1,2 s \ket, \
\widetilde R_- |n_r,0 \ket \propto |n_r - 1,- 2 s \ket.
\end{equation}
As before, these relations follow from the $SU(2)$ algebra which now implies 
that $\widetilde R_\pm$ are raising and lowering operators for 
$\widetilde L = L/2 s$. Hence, by acting with $\widetilde R_\pm$ the eigenvalue
$m$ of $L$ is now shifted by $\pm 2 s$.

One now confirms the value of the Casimir spin $S = n_r + |m|/2 s$ by 
evaluating
\begin{eqnarray}
C |0,m + 2 n_r s \geq 0 \ket&=&(\widetilde L + \widetilde L^2) 
|0,m + 2 n_r s \geq 0 \ket \nonumber \\
&=&\left(\frac{m}{2 s} + n_r\right) \left(\frac{m}{2 s} + n_r + 1\right) 
|0,m + 2 n_r s \geq 0 \ket \nonumber \\
&=&S(S+1) |0,m + 2 n_r s \geq 0 \ket, \nonumber \\
C |0,m - 2 n_r s \leq 0 \ket&=&(- \widetilde L + \widetilde L^2) 
|0,m - 2 n_r s \leq 0 \ket \nonumber \\
&=&\left(- \frac{m}{2 s} + n_r\right)\left(- \frac{m}{2 s} + n_r + 1\right) 
|0,m - 2 n_r s \leq 0 \ket \nonumber \\
&=&S(S+1) |0,m - 2 n_r s \leq 0 \ket.
\end{eqnarray}
The multiplet of degenerate states with the same value of $S$ is again obtained
by repeated applications by $\widetilde R_+$ or $\widetilde R_-$.

As in the case of the $1/r$ potential, for $s \neq 1$ different types of 
unusual multiplets arise. Again, even for integer or half-integer 
$S = n_r + |m|/2 s$, the 
degeneracy of the physical multiplet is not $2 S + 1$ because $m \pm 2 n s$ 
may not be an integer in which case the corresponding wave function is not 
$2 \pi$-periodic. When $S = n_r + |m|/2 s$ is neither an integer nor a 
half-integer, there is again an infinite number of degenerate solutions of the 
Schr\"odinger equation. However, once more, only a finite number of them obeys 
the boundary condition of eq.(\ref{boundary}) and thus belongs to 
${\cal D}[H]$. A sequence of physical and unphysical wave functions is 
illustrated in figure 7.
\begin{figure}[htb]
\begin{center}
\epsfig{file=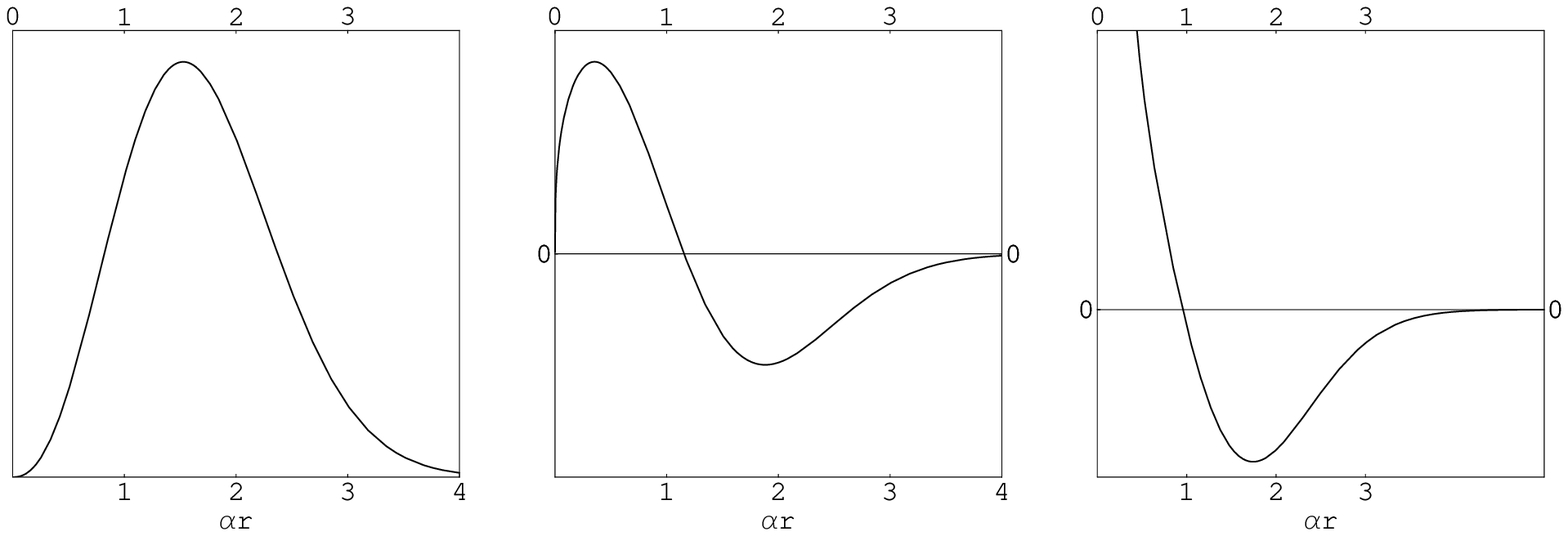,width=15cm}
\end{center}
\caption{\it A sequence of wave functions for the $r^2$ potential with $s = 3$
obtained from repeated applications of $\widetilde R_-$. The quantum numbers
are $n_r = 0$, $m = 7$ (left), $n_r = 1$, $m = 7 - 2s = 1$ (middle), and
$n_r = 2$, $m = 7 - 4s = -5$ (right). The third state in the sequence is 
outside the domain of the Hamiltonian because the corresponding wave function
does not obey the boundary condition of eq.(\ref{boundary}) and the state is 
thus unphysical.}
\end{figure}

\section{Conclusions}

We have considered the physics of a particle confined to the surface of a cone
with deficit angle $\delta$ and bound to its tip by a $1/r$ or an $r^2$ 
potential. In both cases, for rational $s = 1 - \delta/2 \pi$, all
bound classical orbits are closed and there are accidental degeneracies in the
discrete energy spectrum of the quantum system. There is an accidental $SU(2)$
symmetry generated by the Runge-Lenz vector and by the angular momentum.
However, the Runge-Lenz vector is not necessarily a physical operator. For 
example, by acting with the Runge-Lenz vector on a physical state one may
generate an unphysical wave function outside the domain of the Hamiltonian. As 
a result, the representations of the accidental $SU(2)$ symmetry are larger 
than the multiplets of degenerate physical states. In particular, some physical
states are contained in multiplets with an unusual value of the Casimir spin 
$S$ which is neither an integer nor a half-integer. Still, the fractional value
of the spin yields the correct value of the quantized energy.

The particle on a cone provides us with an interesting physical system in which
symmetries manifest themselves in a very unusual manner. Although the 
Hamiltonian commutes with the generators of an $SU(2)$ symmetry, the multiplets
of degenerate states do not always correspond to integer or half-integer 
Casimir spin. This is because the application of the symmetry generators may 
lead us out of the domain of the Hamiltonian. Only the states with 
square-integrable single-valued $2 \pi$-periodic wave functions belong to the
physical spectrum, and all other members of the corresponding ``$SU(2)$'' 
representation must be discarded. Mathematically speaking, the symmetry
generators --- although Hermitean in their respective domain --- do not act as
Hermitean operators in the domain of the Hamiltonian. 

In contrast to many other quantum mechanics problems, in order to understand 
motion on a cone it was necessary to address mathematical issues such as the 
domains of operators as well as Hermiticity versus self-adjointness. Still, we 
have not elaborated on some questions related to different possible 
self-adjoint extensions of the Hamiltonian. For the particle on the cone, such 
issues seem worth investigating. In this work, we have limited ourselves to the
standard Friedrichs extension of the Hamiltonian. Alternative self-adjoint 
extensions correspond to an additional $\delta$-function potential located at 
the tip of the cone. This will modify the problem in an interesting way. In 
particular, we expect that, in the presence of an additional $\delta$-function
potential, the accidental degeneracy will be partly lifted. However, since the 
$\delta$-function only affects states with $m = 0$, some accidental degeneracy 
will remain. The particle on the cone provides us with another example for the 
deep connection between the closedness of all bound classical orbits and 
accidental degeneracies in the discrete spectrum of the Hamiltonian. Even if 
the classical system has various quantum analogues (because there are different
possible self-adjoint extensions) some accidental degeneracy still persists. It
is also remarkable that, like in other cases with accidental symmetries, for 
the particle on the cone semi-classical Bohr-Sommerfeld quantization provides 
the exact quantum energy spectrum.

We are unaware of another system for which a similarly unusual symmetry 
behavior has been observed. It is interesting to ask if symmetry can 
manifest itself in this unusual manner also in other quantum systems. For 
example, cones of graphene may provide a motivation to study accidental 
degeneracies of the Dirac equation on a cone. Also higher-dimensional spaces 
with conical singularities may be worth investigating. In any case, we hope 
that we have convinced the reader that motion on a cone provides an 
illuminating example for a rather unusual manifestation of symmetry in quantum 
mechanics.

\section*{Acknowledgements}

We are indebted to F.\ Niedermayer and C.\ Tretter for illuminating 
discussions. This work is supported in parts by the Schweizerischer 
Nationalfonds.

\end{document}